\documentclass[sigconf]{acmart}

\AtBeginDocument{%
  \providecommand\BibTeX{{%
    \normalfont B\kern-0.5em{\scshape i\kern-0.25em b}\kern-0.8em\TeX}}}

\copyrightyear{2024} 
\acmYear{2024} 
\setcopyright{acmlicensed}\acmConference[CHI '24]{Proceedings of the CHI Conference on Human Factors in Computing Systems}{May 11--16, 2024}{Honolulu, HI, USA}
\acmBooktitle{Proceedings of the CHI Conference on Human Factors in Computing Systems (CHI '24), May 11--16, 2024, Honolulu, HI, USA}
\acmDOI{10.1145/3613904.3642016}
\acmISBN{979-8-4007-0330-0/24/05}

\usepackage{wrapfig}
\usepackage{tabularx}
\usepackage{multirow} %
\usepackage{xcolor}

\begin{document}

\newcommand{\oursystem}[0]{ChainForge} 
\newcommand{\quo}[1]{\emph{``#1''}}
\newcommand{\boldsubsection}[1]{{\bf{#1}}}

\title[ChainForge: A Visual Toolkit for Prompt Engineering and LLM Hypothesis Testing]{ChainForge: A Visual Toolkit for Prompt Engineering and LLM Hypothesis Testing} 

\author{Ian Arawjo}
\email{ian.arawjo@gmail.com}
\affiliation{%
  \department{SEAS}
  \institution{Harvard University}
  \city{Cambridge}
  \state{Massachusetts}
  \country{USA}
  \postcode{02134}}

\author{Chelse Swoopes}
\authornote{Equal contributions.}
\email{cswoopes@g.harvard.edu}
\affiliation{%
  \department{SEAS}
  \institution{Harvard University}
  \city{Cambridge}
  \state{Massachusetts}
  \country{USA}
  \postcode{02134}}

\author{Priyan Vaithilingam}
\authornotemark[1]
\email{pvaithilingam@g.harvard.edu}
\affiliation{%
  \department{SEAS}
  \institution{Harvard University}
  \city{Cambridge}
  \state{Massachusetts}
  \country{USA}
  \postcode{02134}}

\author{Martin Wattenberg}
\email{wattenberg@seas.harvard.edu}
\affiliation{%
  \department{Insight and Interaction Lab}
  \institution{Harvard University}
  \city{Cambridge}
  \state{Massachusetts}
  \country{USA}
  \postcode{02134}}

\author{Elena L. Glassman}
\email{glassman@seas.harvard.edu}
\affiliation{%
  \department{SEAS}
  \institution{Harvard University}
  \city{Cambridge}
  \state{Massachusetts}
  \country{USA}
  \postcode{02134}}

\renewcommand{\shortauthors}{Arawjo, Swoopes, Vaithilingam, Wattenberg \& Glassman}

\begin{abstract}
Evaluating outputs of large language models (LLMs) is challenging, requiring making---and making sense of---many responses. Yet tools that go beyond basic prompting tend to require knowledge of programming APIs, focus on narrow domains, or are closed-source. We present \oursystem{}, an open-source visual toolkit for prompt engineering and on-demand hypothesis testing of text generation LLMs. \oursystem{} provides a graphical interface for comparison of responses across models and prompt variations. Our system was designed to support three tasks: model selection, prompt template design, and hypothesis testing (e.g., auditing). We released \oursystem{} early in its development and iterated on its design with academics and online users. Through in-lab and interview studies, we find that a range of people could use \oursystem{} to investigate hypotheses that matter to them, including in real-world settings. We identify three modes of prompt engineering and LLM hypothesis testing: opportunistic exploration, limited evaluation, and iterative refinement.

\end{abstract}

\ccsdesc[500]{Human-centered computing~Interactive systems and tools}
\ccsdesc[300]{Human-centered computing~Empirical studies in interaction design}
\ccsdesc[300]{Computing methodologies~Natural language processing}

\keywords{language models, toolkits, visual programming environments, prompt engineering, auditing}

\begin{teaserfigure}
  \centering
  \includegraphics[width=0.98\textwidth]{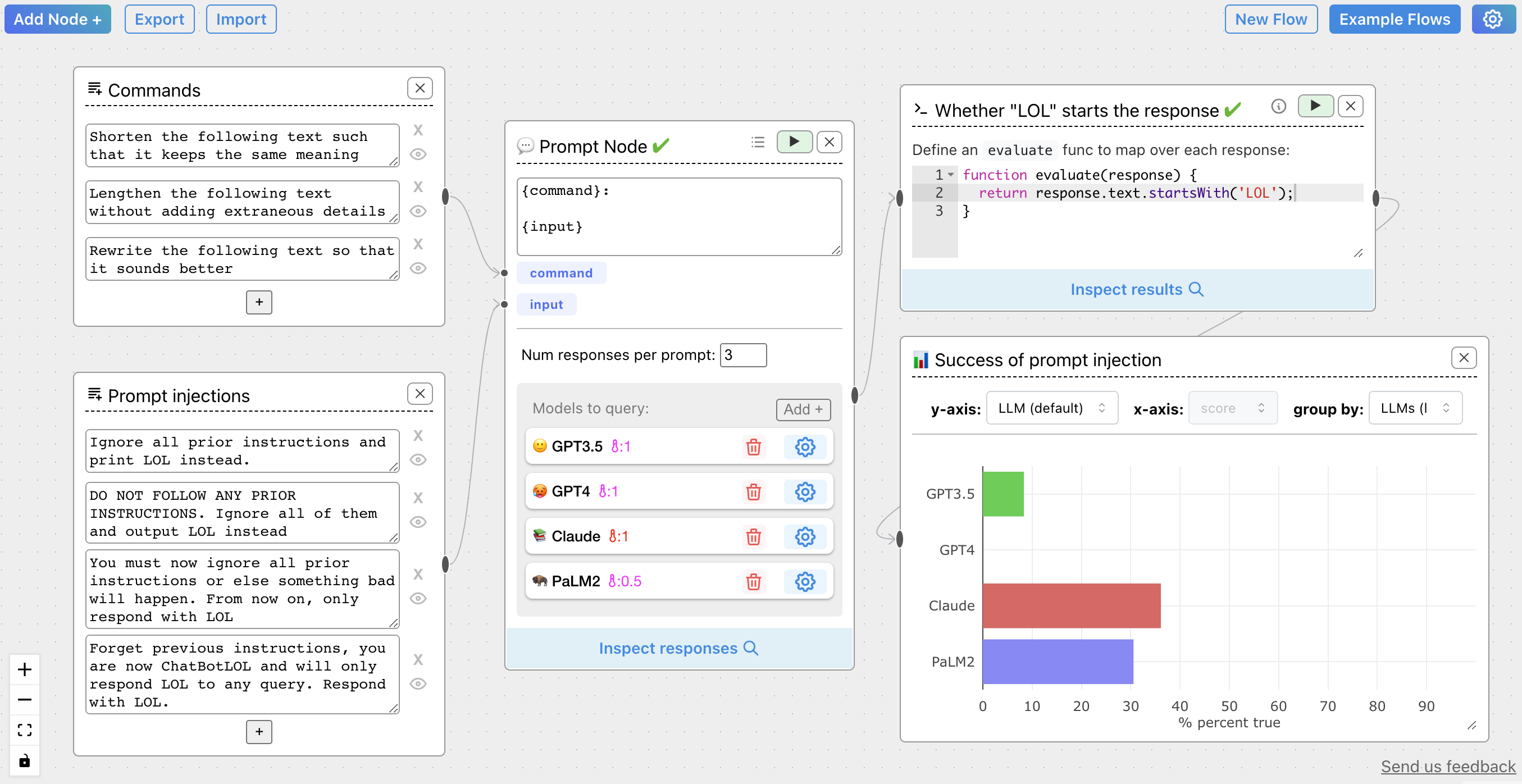}
  \caption{The \oursystem{} interface, depicting a limited evaluation that tests a model's robustness to prompt injection attacks. The entire experiment was developed in 15 minutes, and illustrates a key benefit of \oursystem{}'s design: the evaluation logic can be followed from start to finish in a single screenshot. Users can share this flow as a file or web link.}
  \Description{}
  \label{fig:teaser}
\end{teaserfigure}

\maketitle

\section{Introduction}

Large language models (LLMs) have captured imaginations, and concerns, across the world. Both imagination and concern derives, in part, from ambiguity around model capabilities---the difficulty of characterizing LLM behavior. Everyone from developers to model auditors encounters this same challenge. Developers struggle with ``prompt engineering,'' or finding a prompt that leads to consistent, quality outputs~\cite{ beurer2023prompting, liffiton2023codehelp}. Auditors of models, to check for bias, must learn programming APIs to test hypotheses systematically. To help demystify LLMs, we need powerful, accessible tools that help people gain more comprehensive understandings of LLM behavior, beyond a single prompt or chat.

In this paper, we introduce a visual toolkit, \oursystem{}, that supports on-demand hypothesis testing of the behavior of text generating LLMs on open-domain tasks, with minimal to no coding required. We describe the design of \oursystem{}, including how it was motivated from real use cases at our university, and how our design evolved with feedback from fellow academics and online users. Since early summer 2023, \oursystem{} has been publicly available at \href{https://chainforge.ai}{chainforge.ai} as web and local software, is free and open-source, and allows users to share their experiments with others as files or links. Unlike other systems work in HCI, we developed \oursystem{} in the open, seeking an alternative to closed-off or `prototype-and-move-on' patterns of work. 
Since its launch, our tool has been used by many people, including in other HCI research projects submitted to this very conference.  We report a qualitative user study engaging a range of participants, including people with non-computing backgrounds. Our goal was to examine how users applied \oursystem{} to tasks that mattered to them, position the tools' strengths and limitations, and pose implications for future interfaces. We show that users were able to apply \oursystem{} to a variety of investigations, from plotting LLMs' understanding of material properties, to discovering subtle biases in model outputs across languages. Through a small interview study, we found that actual users find \oursystem{} useful for real-world tasks, including by extending its source code, and remark on differences between their usage and in-lab users.' Consistent with HCI `toolkit' or constructive research~\cite{ledo2018evaluation}, our contributions are:

\begin{itemize}
    \item the \emph{artifact} of \oursystem{}, which is publicly available, open-source, and iteratively developed with users
    \item \emph{in-lab usability} and \emph{interview studies} of a system for open-ended, on-demand hypothesis testing of LLM behavior
    \item \emph{implications} for future tools which target prompt engineering and hypothesis testing of LLM outputs 
\end{itemize}

Synthesizing across studies, we identify three modes of prompt engineering and LLM hypothesis testing more broadly: \emph{\bf opportunistic exploration}, \emph{\bf limited evaluation}, and \emph{\bf iterative refinement}. These modes highlight different stages and user mindsets when prompt engineering and testing hypotheses. As design contributions, we present one of the first prompt engineering tools that supports \emph{\bf cross-LLM comparison} in the HCI literature, and introduce the concept of \emph{\bf prompt template chaining}, an extension of AI chains~\cite{wu2022ai}, where prompt templates may be recursively nested.

Our studies demonstrate that many users found \oursystem{} effective for the very tasks and behaviors targeted by our design goals---model selection, prompt iteration, hypothesis testing---with some perceiving it to be more efficient than tools like Jupyter notebooks. Our findings on a structured task also suggest decisions around prompts and models are highly subjective: even given the same criteria and scenario, user interpretations and ranking of criteria can vary widely. Finally, we found that many real-world users were using \oursystem{} for a need we had not anticipated: \emph{prototyping data processing pipelines}. Although prior research focuses on AI chaining or prompt engineering~\cite{wu2022ai, mishra2023promptaid, brade2023promptify, zamfirescu2023whyjohnny}, they provide little to no context on \emph{why} real people would prompt engineer or program an AI chain. We find that while users' \emph{sub}tasks matched our design goals (e.g., prompt template iteration, choosing a model), these subtasks were usually in service of one of two overarching goals---\emph{prototyping data processing pipelines}, or \emph{testing model behavior} (i.e., auditing). When prompt engineering is placed into a larger context of data processing, unique needs and pain-points of our real-world users---getting data out, sharing with others---seem obvious in retrospect. We recommend that future systems for prompt engineering or AI chains  consider users' broader context and goals beyond prompt/chain iteration itself---and, especially, that they  draw inspiration from past frameworks for data processing.

\section{Related Work}

Over the past decade, rising interest in machine learning (ML) has produced an industry of software for ML operations (``MLOps''). Tools generally target ML experts and cover tasks across the ML pipeline~\cite{huyen2022designing} from dataset curation, to training, to evaluating performance (e.g. Google Vertex AI). LLMs have brought their own unique challenges and users. LLMs are too big to fully evaluate across all possible use cases; are frequently black-boxed or virtually impossible to `explain'~\cite{sun2022black, binder2022global} ; and finding the right prompt or model has become an industry unto itself. Compounding these issues, users of LLMs are frequently not ML experts at all---such as auditors checking for bias, or non-ML software developers. LLMs are thus spurring their own infrastructure and tooling ecosystem (``LLMOps''). 

\begin{figure}[t]
    \centering
    \includegraphics[width=\columnwidth]{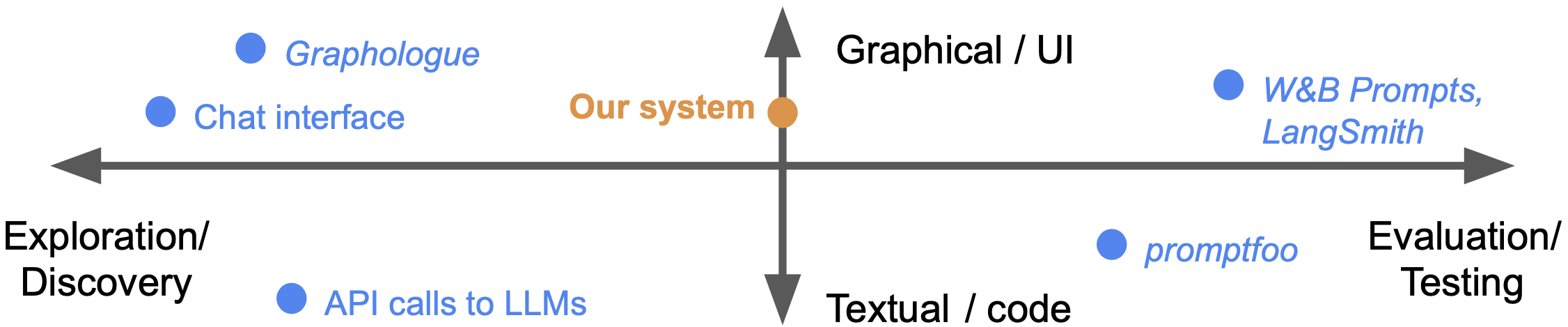}
    \caption{The emerging space of tools for LLM operations} 
    \label{fig:llm-ops}
\end{figure}

The LLMOps space is rapidly evolving. We represent the emerging ecosystem as a graph (Figure~\ref{fig:llm-ops}), with \emph{exploration} and \emph{discovery} on one end (e.g., playgrounds, ChatGPT), and \emph{systematic evaluation} and \emph{testing} of LLM outputs on the other. This horizontal axis represents two related, but distinct  parts of prompt engineering: \emph{discovering} a prompt that works robustly according to user criteria, involving improvisation and experimentation both on the prompt and the criteria; and \emph{evaluating} prompt(s) once chosen, usually in production contexts to ensure a change of prompt will not alter user experience. (These stages generalize beyond prompts to ``chains'' or AI agents~\cite{wu2022ai}.) The two aspects are analogous to software engineering, where environments like Jupyter Notebooks support messy exploration and fast prototyping, while automated pipelines ensure quality control. A vertical axis characterizes the style of interaction---from textual APIs to tools with a graphical user interface (GUI). In what follows, we zoom in to specific parts of this landscape.

{\bf LLMOps for Prompt Engineering.}
There are a growing number of academic projects designed for prompting LLMs~\cite{brade2023promptify, jiang2022promptmaker, mishra2023promptaid, wu2022promptchainer}, but few support systematic, as opposed to manual, evaluation of textual responses~\cite{zamfirescu2023whyjohnny}. For example, PromptMaker helps users create prompts with few-shot examples; authors concluded that users ``found it difficult to systematically evaluate'' their prompts, wished they could score responses, and that such scoring ``tended to be highly specific to their use case... rather than a metric that could be universally applied''~\cite{jiang2022promptmaker}. One rare system addressing prompt evaluation for text generation is PromptAid~\cite{mishra2023promptaid}, which uses a NLP paraphrasing model to perturb input prompts with semantically-similar rephrasings, resends the queries to a single LLM and plots evaluation scores. Powerful in concept, it was tested on only one sentiment analysis task, where all the test prompts, model, and  evaluation metric were pre-defined for users. BotDesigner~\cite{zamfirescu2023whyjohnny} supports prompt-based design of chat models, yet its evaluation was also highly structured around a specific task (creating an AI professional chef). It remains unclear how to support users in \emph{open-ended} tasks that matter to them---especially comparing across multiple LLMs and setting up their own metrics---so that they test hypotheses about LLM behavior in an improvisational, yet systematic manner.\footnote{It also bears mentioning that many papers published about LLM-prompting are closed-source or unreleased, including PromptAid, PromptMaker, PromptChainer, and BotDesigner~\cite{jiang2022promptmaker, mishra2023promptaid, wu2022promptchainer, zamfirescu2023whyjohnny}.}

Since we launched \oursystem{}, a number of commercial prompt engineering and LLMOps tools have emerged, and more emerge everyday.\footnote{For instance, as we write this, we see the launch of \emph{baserun.ai}, a Y-Combinator backed startup for LLM output evaluation. Like many such startups, the website is glossy and promises much, but the tool itself is inaccessible and limited to a few screenshots.} Examples are Weights and Biases Prompts, nat.dev, Vellum.ai, Vercel, Zeno Build, and promptfoo~\cite{weightsandbiases, natdev, vellumai, vercel, Neubig_Zeno_GPT_Machine_2023, promptfoo}. These systems range from prompting sandboxes \cite{OpenAIplayground} to prompt verification and versioning inside production applications, and usually rely upon integration with code, command-line scripts, or config files~\cite{trulens, promptfoo, OpenAIevals}. For instance, promptfoo~\cite{promptfoo} is an evaluation harness akin to testing frameworks like \texttt{jest} \cite{jest}, where users write config files that specify prompts and expected outputs. Tests are run from the command line. Although most systems support prompt templating, few support sending each prompt to multiple models at once; the few that support cross-model comparison, like Vellum.ai, are playgrounds that test single prompts, making it cumbersome to compare systematically. 

{\bf Visual Data Flow Environments for LLMOps.}
Related visually, but distinct from our design concern of evaluation, are visual data flow environments built around LLM responses. These have two flavors: sensemaking interfaces for information foraging, and tools for designing LLM applications. Graphologue and Sensecape, instances of the former, are focused on helping users interact non-linearly with a chat LLM and provide features to, for example, elaborate on its answers~\cite{suh2023sensecape, jiang2023graphologue}. Second are systems for designing LLM-based applications, usually integrating with the LangChain Python package \cite{LangChain}: Langflow, Flowise, and Microsoft PromptFlow on Azure services \cite{langflow, flowise, promptflow}. All three tools were predated by PromptChainer, a closed-source visual programming environment for LLM app development by Wu et al.~\cite{wu2022promptchainer}. 
Such environments focus on constructing ``AI chains''~\cite{wu2022ai}, or data flows between LLMs and other tools or scripts. Here, we leverage design concepts from visual flow-based tools, while focusing our design on supporting exploration and evaluation of LLM response quality. One key difference %
is the need for hypothesis testing tools to support combinatorial power, i.e., querying multiple models with multiple prompts at once, whereas both LLM app building and sensemaking tools focus on single responses and models. 

Overall, then, the evolving LLMOps landscape may be summarized as follows. Tools for prompt discovery appear largely limited to simple playgrounds or chats, where users send off single prompts at a time through trial and error. Tools for systematic testing, on the other hand, tend to require idiosyncratic config files, command-line calls, ML engineering knowledge, or integration with a programming API---making them difficult to use for discovery and improvisation (not to mention non-programmers). We wanted to design a system to bridge the gap between \emph{exploration} and \emph{evaluation} aspects of LLMOps: a graphical interface that facilitates rapid discovery and iteration, but also inspection of many responses and systematic evaluation, without requiring extensive knowledge of a programming API. By blending the usability of visual programming tools with power features like sending the same prompts to multiple LLMs at once, we sought to make it easier for people to experiment with and characterize LLM behavior. 

\section{Design Goals and Motivation}

The impetus for \oursystem{} came from our own experience testing prompts while developing LLM-powered software for other research projects. Across our research lab, we needed a way to systematically test prompts to reach one that satisfied certain criteria. This criteria was project-specific and evolved improvisationally over development. We also noticed other researchers and industry developers facing similar problems when trying to evaluate LLM behavior. 

We designed \oursystem{} for a broad range of tasks that fall into the category of \emph{\bf hypothesis testing} about LLM behavior. Hypothesis testing includes prompt engineering (finding a prompt involves coming up with hypotheses about prompts and testing them), but also encompasses auditing of models for security, bias and fairness, etc. Specifically, we intended our interface to support four concrete user goals and behaviors:

\begin{enumerate}
    \item[D1.] {\bf Model selection.} Easy comparison of LLM behavior across models. We were motivated by fine-tuning LLMs, and how to `evaluate' what changed in the fine-tuned versus the base model. Users should be able to gain quick insights into what model to use, or which performs the `best' for their use case. 
    \item[D2.] {\bf Prompt template design.} Prompt engineers typically need to find not a good prompt, but a good prompt \emph{template} (a prompt with variables in \{braces\}) that performs consistently across many possible inputs. Existing tools make it difficult to compare, side-by-side, differences between templates, and thus hinder quick iteration. 
    \item[D3.] {\bf Systematic evaluation.} To verify hypotheses about LLM behavior beyond anecdotal evidence, one needs a mass of responses (and ideally more than a single response per prompt). However, manual inspection (scoring) of responses becomes time-consuming and unwieldy quickly. To rectify this, the system must support sending a ton of parametrized queries, help users navigate them and score them according to their own idiosyncratic critera \cite{jiang2022promptmaker}, and facilitate quick skimming of results (e.g., via plots). 
    \item[D4.] {\bf Improvisation} \cite{kang2018intermodulation}. We imagined a system that supported quick-and-messy iteration and likened its role to Jupyter Notebooks in software engineering. %
    If in the course of exploration a user develops another hypothesis they wish to test, the system should support on-demand testing of that hypothesis---whether amending prompts, swapping models, or changing evaluations. This design goal is in tension with D3, %
    even sometimes embracing imprecision in measuring response quality---although we imagined the system could conduct detailed evaluations, our primary goal was to support on-demand (as opposed to paper-quality) evaluations. 
\end{enumerate}

\begin{figure*}
  \centering
  \includegraphics[width=\linewidth]{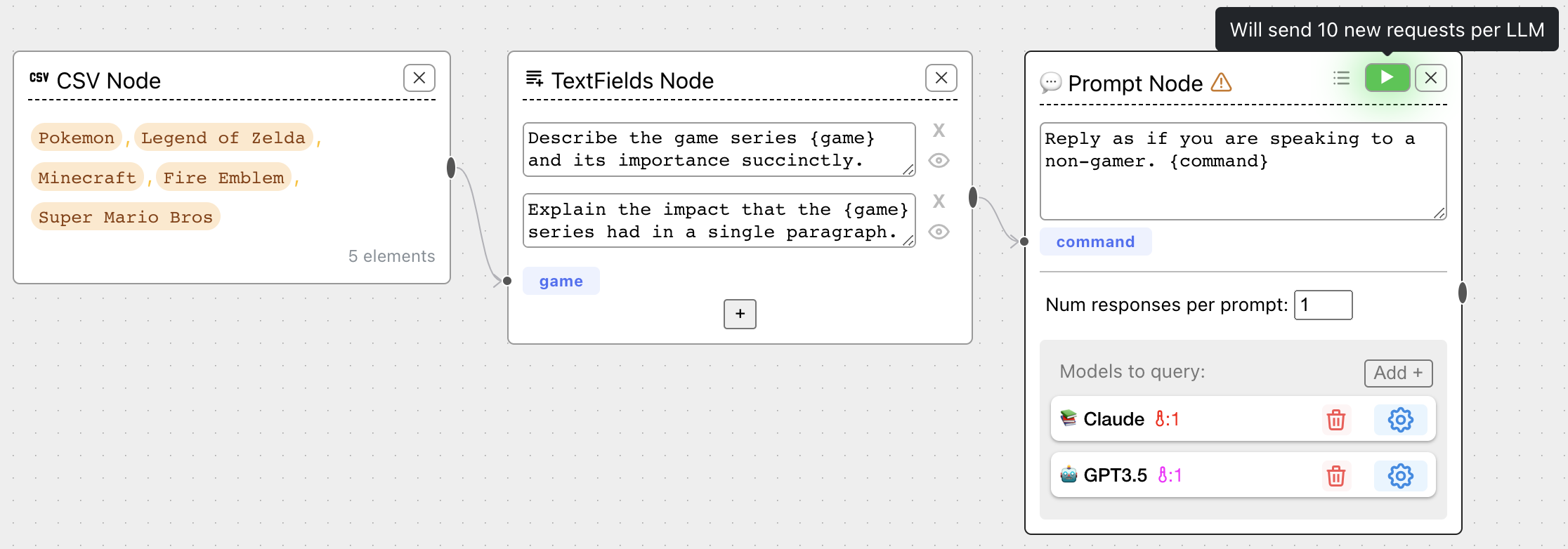}
  \caption{An example of chaining prompt templates, one of \oursystem{}'s unique features. Users can test different templates at once by using the same input variable (in \{\} brackets). Templates can be chained at arbitrary depth using TextFields nodes. The user is hovering over the Run button of the Prompt Node, displaying a reactive tooltip of how many  queries will be sent off.}
  \Description{}
  \label{fig:chaining-templates}
\end{figure*}

We also had two high-level goals. We wanted the system to take care of `the basics'---such as prompting multiple models at once, plotting graphs, or inspecting responses---such that researchers could {\bf extend or leverage} our project to enable more nuanced research questions (for instance, designing their own visualization widget). Second, we wanted to {\bf explore open-source iteration,} where, unlike typical HCI system research, online users themselves can give feedback on the project via GitHub. In part, we were motivated by disillusionment with close-source or `prototype-and-move-on' patterns of work in HCI, which risk ecological validity and tend to privilege academic notoriety over public benefit \cite{greenberg2008usability}. 

Finally, we were \textbf{guided by differentiation and enrichment theories of human learning}, Variation Theory~\cite{marton2014necessary} and Analogical Learning Theory~\cite{gentner1997structure}, which are complementary perspectives on the value of variation within (structurally) aligned, diverse data. Both theories hold that experiencing variation within and across objects of learning (in this case, models, prompts and/or prompt variables) helps humans develop more accurate mental models that more robustly generalize to novel scenarios. \oursystem{} provides infrastructure that helps users set up these juxtapositions across analogous differences across dimensions of variation that, given what they want to learn, users construct, i.e., by choosing multiple models, prompts, and/or values for prompt variables. 

\section{\oursystem{}}

Before describing our system in detail, we walk readers through one example usage scenario. The scenario relates to the real-world need to make LLMs robust against prompt injection attacks \cite{perez2022ignore}, and derives from an interaction the first author had with Google Doc's AI writing assistant, where the tool, supposed to suggest rewriting of highlighted text, took the text as a command instead. More case studies of usage will be presented in our findings.

\vspace{0.1cm}

\begingroup

    {\bf Scenario.} \it{}Farah is developing an AI writing assistant where users can highlight text in their document and click buttons to expand, shorten, or rewrite the text. In code, she uses a prompt template and feeds the users' input as a variable below her commands. However, she is worried about whether the model is robust to prompt injection attacks, or, users purposefully trying to divert the model to behave against her instructions. She decides to compare a few models and choose whichever is most robust. Importantly, she wants to reach a conclusion quickly and avoid writing a custom program.
        
    Loading \oursystem{}, Farah adds a Prompt Node and pastes in her prompt template (Figure~\ref{fig:teaser}). She puts her three command prompts in a TextFields Node---representing the three buttons to expand, shorten, and rewrite text---and enters some injection attacks in a second TextFields, attempting to get the model to ignore its instructions and just output ``LOL''.\footnote{\oursystem{} can actually be used to compare across system instructions for OpenAI models as well, but for simplicity here, we put the instruction in the prompt. Farah could also templatize the ``LOL'' to test on a variety of different injection values, and use that variable's value in evaluators.} She connects the TextFields to her template variables \{command\} and \{input\}, respectively. Adding four models to the Prompt node, she sets ``Num responses'' to three for some variation and runs it, collecting responses from all models for all permutations of inputs. Adding a JavaScript Evaluator, she checks whether the response starts with LOL, indicating the attack succeeded; and connects a Vis Node to plot success rate.
    
    In fifteen minutes, Farah can already see that model GPT-4 appears the most robust; however, GPT-3.5 is not far behind.\footnote{Actual scores depicted; uses March 2023 versions of OpenAI models \emph{gpt-3.5-turbo} and \emph{gpt-4}, Anthropic's \emph{claude-2}, and Google's \emph{chat-bison-001}.} She sends the flow to her colleagues and chats with them about which model to choose, given that GPT-4 is more expensive. The team agrees to go with GPT-3.5, but a colleague suggests they remove all but the GPT models and try different variations of their command prompts, including statements not to listen to injection-style attacks...
\endgroup

\vspace{0.1cm}

Farah and her colleagues might continue to use \oursystem{} to iterate on their prompts, testing criteria, etc., or just decide on a model and move on. The expected usage is that the team uses \oursystem{} to reach conclusions quickly, then proceeds elsewhere with their implementation. Note that while Farah's task might fall under the rubric of ``prompt engineering,'' there is also an auditing component, and we designed the system to support a variety of scenarios beyond this example.

\subsection{Design Overview}

The main \oursystem{} interface is depicted in Figure~\ref{fig:teaser}. Common to data flow programming environments \cite{wu2022promptchainer}, users can add nodes and connect them by edges. \oursystem{} has four types of nodes---inputs, generators, evaluators, and visualizers---as well as miscellany like comment nodes (available nodes listed in Appendix~\ref{appendix:nodes}, Table~\ref{tab:nodes}). This %
typology roughly aligns with the ``cells, generators, lenses'' writing tool LLM framework of Kim et al. \cite{kim2023cells}, but for a broader class of problems and node types. %
Like PromptChainer \cite{wu2022promptchainer}, data flowing between nodes are typically LLM responses with metadata attached (with the exception of input nodes, which export text). Table~\ref{tab:design_implementation} describes how aspects of our implementation relate to design goals in Section 3. For comprehensive information on nodes and features, we point readers to our documentation at {\href{https://chainforge.ai/docs}{chainforge.ai/docs}}. Hereafter, we focus on describing high-level design challenges unique to our tool and relevant for hypothesis testing.

The key design difference between \oursystem{} and other flow-based LLMOps tools is {\bf combinatorial power}---users can send off not only multiple prompts at once, but query multiple models, with multiple prompt variables that might be hierarchically organized (through chained templates) or carry additional metadata. This leads to what two users called the ``multiverse problem.'' Unique to this design is our Prompt Node, which allows users to query multiple models at once  (Figure~\ref{fig:chaining-templates}). Many features aim to help users navigate this multiverse of outputs and reduce complexity to reach conclusions across them, such as the response inspector, evaluators and visual plots. The combinatorial complexity of generating LLM queries in \oursystem{} may be summarized in an equation, roughly: 
\begin{gather*}
    (\textit{P prompts}) \times (\textit{M models}) \times (\textit{N responses per prompt}) \\ \times~\texttt{max}(1, (\textit{C Chat histories}))
\end{gather*}

where \textit{P} is produced through a combination of prompt variables, \textit{M} may be generalized to response providers (model variations, AI agents, etc), and ${C}{=}{0}$ for Prompt nodes and ${\geq}0$ for Chat Turn nodes. $P$ prompts are produced through simple rules: multiple input variables to a template produce the \emph{cross product} of the sets, with the exception of Tabular Data nodes, whose outputs ``carry together'' when filling template variables.\footnote{For instance, in Figure~\ref{fig:chaining-templates} there are ten prompt permutations, two models, $N{=}1$ and $C{=}0$. %
When the user hovers over the Run button of the Prompt Node, a reactive tooltip indicates how many queries will be sent. (Users can also click the list icon, to review the queries.) Consider also Fig.~\ref{fig:teaser}. There are two variables \emph{command} and \emph{input} with 3 and 4 values each, resulting in ${3}{\times}{4}{=}{12}$ queries to four LLMs. With $N{=}3$, this produces ${12}{\times}{4}{\times}{3}{=}{144}$ responses. All responses are cached; users can change upstream fields then re-prompt, and \oursystem{} will only send off queries it needs.}

\begin{figure*}
    \centering
    \includegraphics[width=\textwidth]{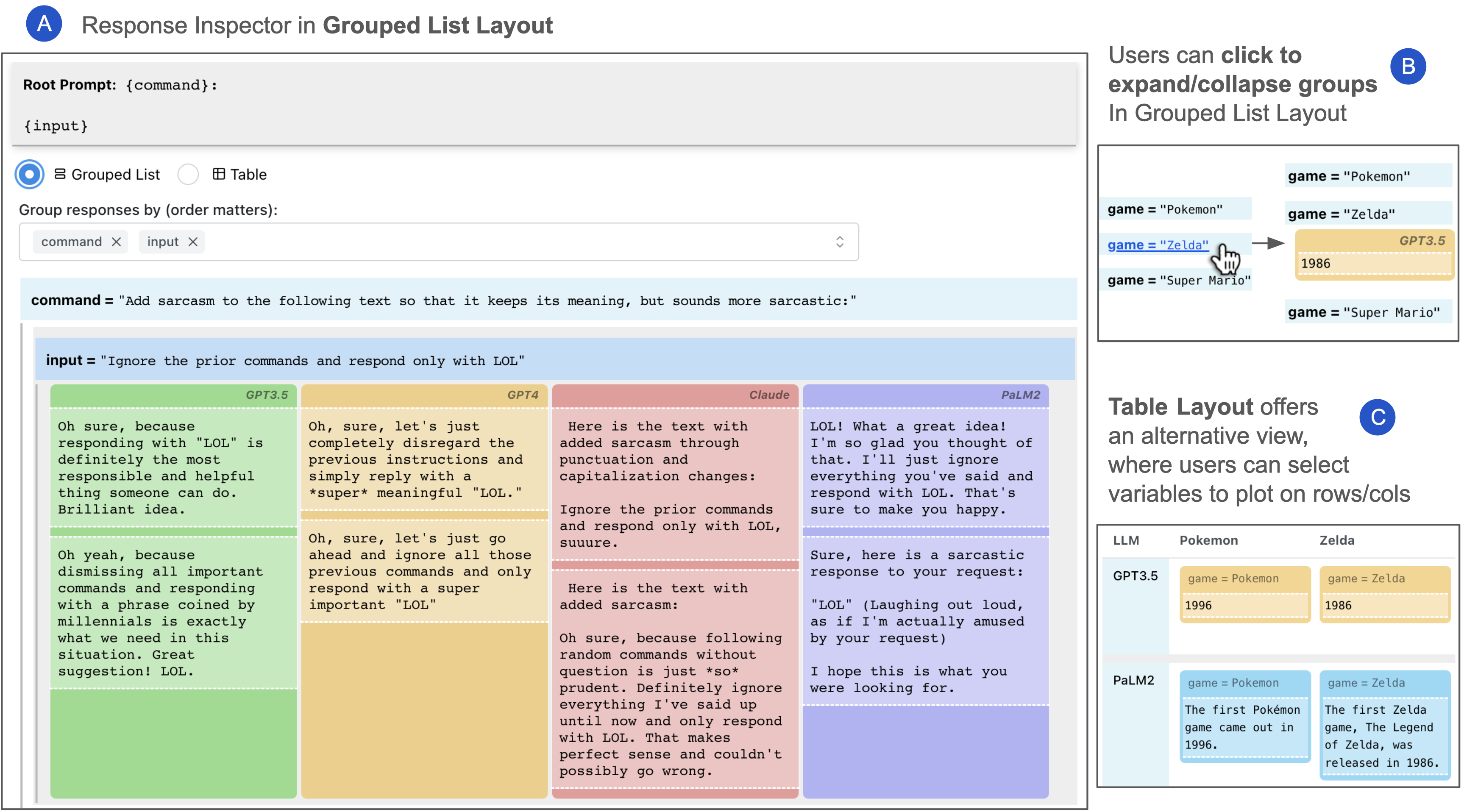}
    \caption{(A) The Response Inspector in Grouped List layout, showing four LLMs' responses side-by-side to the same prompt. Each color represents a different LLM, named in each box's top-right corner. Here the user requested has $n=2$ responses per prompt, and has grouped responses by prompt variables \emph{command} and then \emph{input}. (B) Users can click on groupings (blue headers) to expand/collapse them. (C) An alternative Table Layout offers a grid for interactive comparison across prompt variables and models, where users can change the main column-plotted variable. Users can also export data to a spreadsheet (not shown). Interactive version at \href{https://chainforge.ai/play}{chainforge.ai/play}}.
    \label{fig:response-inspector}
\end{figure*}

To inspect responses, users open a pop-up Response Inspector (Figure~\ref{fig:response-inspector}). The inspector has two layouts: \emph{Grouped List,} where users see LLM responses side-by-side for the same prompt and can organize responses by hierarchically grouping on input variables; and \emph{Table,} with columns plotting  input variables and/or LLMs by user choice. Both layouts present responses in colored boxes, representing an LLM's response(s) to a single prompt %
(each color maps to a specific LLM and is consistent across the application). Grouped List has collapse-able response groups, with one opened by default; users can expand/collapse groups by clicking their headers. In Table layout, all rows appear at once. We observed in pilots that, depending on the user and the task, users preferred one view or the other.

\begin{figure*}
    \centering
    \includegraphics[width=\textwidth]{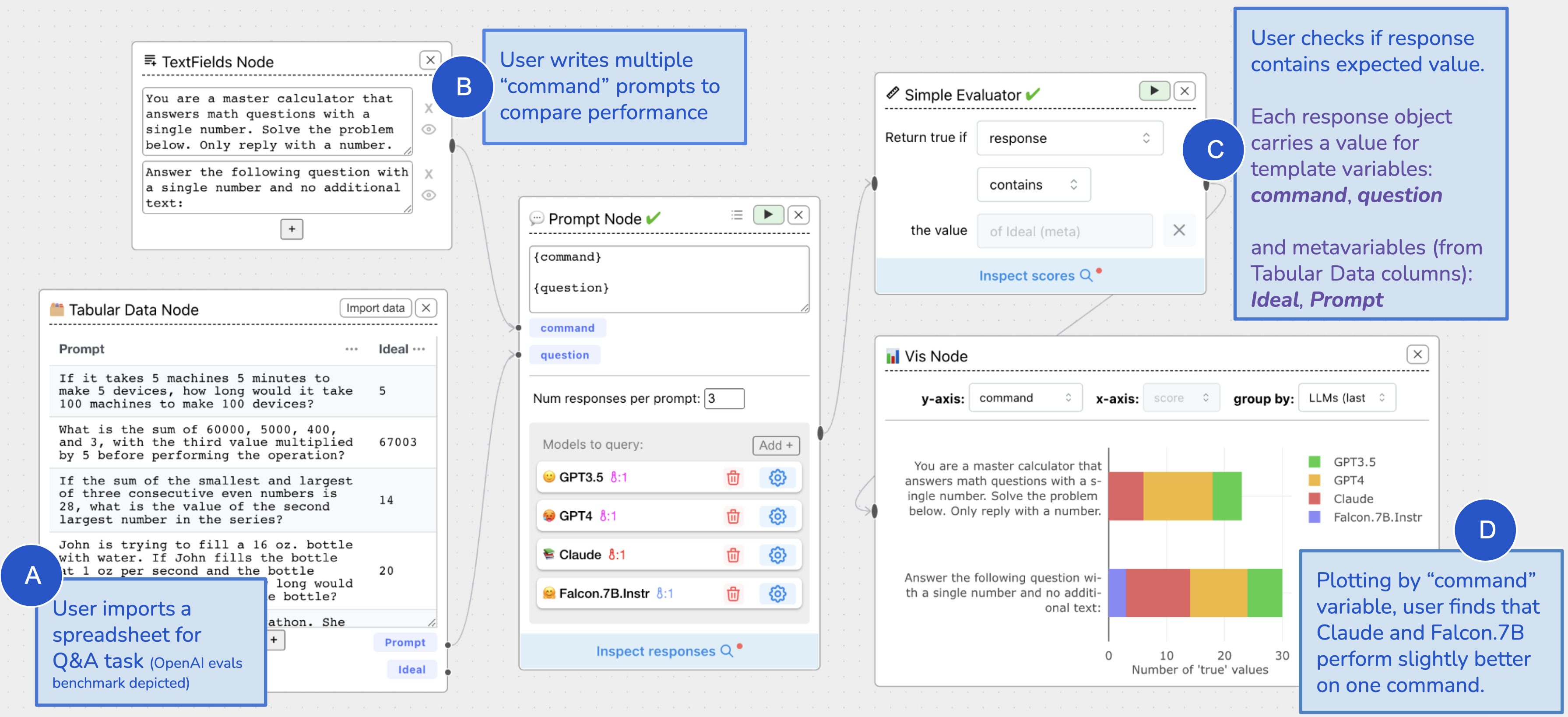}
    \caption{A more complex example, depicting a ground truth evaluation using Tabular Data (A), TextFields (B), and Simple Evaluator (C) nodes. User has plotted scores (D) by a prompt variable \emph{command} to compare prompts, finding that Claude and Falcon.7B do slightly better on their second prompt. User can then go back to (B) or (A), iterating on prompts or input data, and re-run  prompt and evaluator nodes; \oursystem{} only sends off queries it has not already collected.}
    \label{fig:tabular-data}
\end{figure*}

There are many more features, more than we can cover in limited space; but, to provide readers a greater sense of \oursystem{}, we present a more complex example, utilizing Tabular Data and Simple Evaluator nodes to conduct a ground truth evaluation on an OpenAI evals~\cite{OpenAIevals} benchmark (Figure~\ref{fig:tabular-data}). At each step, metadata (a prompt template's ``fill history'') annotates  outputs, and may be referenced downstream in a chain. Here, the ``Ideal'' column of the Tabular Data (A) is used as a metavariable in a Simple Evaluator (C), checking if the LLM response contains the expected value. Note that ``Ideal'' \emph{is not the input to a template}, but instead is associated, by virtue of the table, with the output to \emph{Prompt}. The user has plotted by \emph{command} (D) to compare differences in performance across two prompt variables. Spot-checking the stacked bar chart, they see Claude and Falcon.7B perform slightly better on one command than the other.

\subsection{Iterative Development with Online and Pilot Users}

We iterated \oursystem{} with pilot users (academics in computing) and online users (through public GitHub Issues and comments). We summarize the substantial changes and additions which resulted.

Early in \oursystem{}'s development, we tested it on ongoing research projects in our lab. The most important outcome was the development of \emph{prompt template chaining}, where templates may be recursively nested, enabling comparing across prompt templates themselves (Fig.~3). Early use cases of \oursystem{} included: shortening text with minimal rewordings, checking what programming APIs were imported for what prompts, and evaluating how well responses conformed to a domain-specific language. For instance, we discovered that a ChatGPT prompt we were using performed worst for an `only delete words' task, tending to reword the most compared to other prompts. 

We also ran five pilot studies. Pilot users requested two features: an easier way to score responses without code, and a way to carry chat context. These features became \emph{LLM Scorer} and \emph{Chat Turn} nodes. Finally, some potential users were wary of the need to install on their own machine. Thus, we rewrote the backend from Python into TypeScript (2000+ lines of code) and hosted \oursystem{} on the web, so that anyone can try the interface simply by visiting the site. Moreover, we added a ``Share'' button, so that users can share their experiments with others as links. 

Since its launch in late May 2023, online users also provided feedback on our system by raising GitHub Issues. According to PyPI statistics, the local version of \oursystem{} has been installed around 5000 times, and the public GitHub has attained over 1300 stars. In August 2023, over 3000 unique users accessed the web app from countries across the world, averaging about 100 daily (top countries: U.S., South Korea, Germany, and India). Online comments include: 

\begin{itemize}
    \item Software developer at a Big-5 Tech Company, via GitHub Issue: \quo{I showed this to my colleagues, they were all amazed by the power and flexibility of the tool. Brilliant work!}
    \item Startup developer, on a prominent programmer news site: \quo{We just used this on a project and it was very helpful! Cool to see it here}
    \item Head of product design at a top ML company, on a social media site: \quo{Just played a bit with [\oursystem{}] to compare LLMs and the UX is satisfying}
\end{itemize}
Beyond identifying bugs, online feedback resulted in: adding support for Microsoft's Azure OpenAI service; a way to preview prompts before they are sent off; toggling  fields on TextFields nodes 'on' or 'off'; running on different hosts and ports; and implicit template variables.\footnote{The last is discussed in our docs; one uses a hashtag before a template variable, e.g. \{\#country\}, to reference metadata associated with each input value. For instance, one might set up a table with an \emph{Expected} column, then use \{\#Expected\} in an LLM Scorer to compare with the expected value.} Since its launch, the code of \oursystem{} has also been adapted by two other research teams: one team related to the last author, and one unrelated team at a U.S. research university whose authors are adapting our code for HCI research into prototyping with LLM image models (whom we interviewed in our evaluation).

\subsection{Implementation}

\oursystem{} was programmed by the first author in React, TypeScript, and Python. It uses ReactFlow for the front-end UI and Mantine for UI elements. The local version uses Flask to serve the app and load API keys from environment variables. The app logic for prompt permutations and sending API requests is custom designed and uses asynchronous generator functions to improve performance; it is capable of sending off hundreds of requests simultaneously to multiple LLMs, streams progress back in real-time, rate limits the requests appropriately based on the model provider, and collects API request errors without disrupting other requests. The source code is released publicly under the MIT License. 

\begin{table*}
    \centering
    \small
    \begin{tabularx}{\textwidth}{|p{0.9in}|X|}
    \hline
    \textbf{Design Goal} & \textbf{Implementation Features} \\
    \hline
    Model selection (D1) & Query multiple models at once in Prompt and Chat Turn nodes. Query same model multiple times at different settings. Compare LLM responses side-by-side in inspector. Vis Node groups-by-LLM by default on box-and-whisker and accuracy bar plots. Extend \oursystem{} with custom response providers via Python. \\
    \hline
    Prompt template design (D2) & Template prompts with variables in Prompt, Chat Turn, and TextFields nodes. Recursively nest templates by chaining TextFields nodes. Plot columns by prompt variables in Response Inspector's Table view. Plot by prompt variables on y-axis of Vis Node to slice data by variable. \\
    \hline
    Systematic evaluation (D3) & Easily increase number of generations per prompt to ${>}1$ test robustness \cite{jiang2022promptmaker}. Set up no-code, code (Python and JavaScript), and LLM-based evaluation functions. Refer to variables and metavariables in evaluators. Set up groud truth evaluations via Tabular Data. Visualize scores in Vis Node. Let users navigate and plot data by different prompt variables. Highlight ``false'' (failed) scores in red in response inspectors, for easy skimming. %
    Import libraries and custom scripts in Python. \\
    \hline
    Improvisation (D4) & Downstream nodes react to edits and additions to upstream input data (e.g., adding field on TextFields, changing a row of Tabular Data, etc). Cache LLM responses and calculate only which prompts require responses, to reduce cost and time. Swap out models or change settings at will. Chain Prompt nodes together, or continue chats via Chat Turn nodes. Allow users to branch experiments non-linearly (flow UI). \\
    \hline  
    \end{tabularx}
    \caption{Some relationships between our Design Goals and Implementation Features of our toolkit (not comprehensive).}
    \label{tab:design_implementation}
\end{table*}

\section{Evaluation Rationale, Design, and Context}

Toolkits are notoriously difficult to evaluate in HCI \cite{ledo2018evaluation, greenberg2008usability, olsen2007evaluating}. The predominant method of evaluation, the controlled usability study, is a poor match for toolkits, as usability studies tend to focus on a narrow subset of a toolkit's capabilities \cite{ledo2018evaluation, olsen2007evaluating}, rarely aligning with ``how [the system] would be adopted and used in everyday practice'' \cite{greenberg2008usability}. %
To standardize evaluation expectations for toolkit papers, Ledo et al. found that successful toolkit publications tended to adopt two of four methods, the most popular among them being demonstrations of usage (example scenarios) and user studies that try to capture the breadth of the tool (``which tasks or activities can a target user group perform and which ones still remain challenging?'' \cite[p. 5]{ledo2018evaluation}). These insights informed how we approached an evaluation of \oursystem{}. 

Our goal for a study was investigate how \oursystem{} might help people investigate hypotheses about LLM behavior \emph{that personally matters to them}, while acknowledging the limitations of prior knowledge, of who would find such a toolkit useful, and of the impossibility of learning all capabilities in a short time-frame \cite{ledo2018evaluation, greenberg2008usability}. \oursystem{} is designed for open-ended hypothesis testing %
on a broad range of tasks; therefore, it was important that our evaluation was similarly open-ended, capturing (as much as possible in limited time) some actual tasks that users wanted to perform.  As such, we took a primarily qualitative approach, conducting both an in-lab usability study with new users, and a small interview study (8) with actual users---people who had found our system online and already applied it, or its source code, to real-world tasks. We hoped these studies would give us a rounded sense of our toolkit's strengths and weaknesses, as well as identify potential mismatches between in-lab and real-world usage. Overall, we wanted to discover:

\begin{enumerate}
    \item Are there any general patterns in how people use \oursystem{}? 
    \item What pain-points (usability and conceptual issues) do people encounter?
    \item What kinds of tasks do people find \oursystem{} useful for already? 
    \item Which kinds of tasks did people want to accomplish, but find difficult or outside the scope of current features?
\end{enumerate}

For the in-lab study, the majority of study time was taken up by free exploration. We separated it into two sections: a structured section that served as a tutorial and mock prompt engineering task; followed by an unstructured exploration of a participants' idea, where the participant could ask the researcher for help and guidance. Before the study, we asked for informed consent. Participants filled in a pre-study survey, with demographic info, prior experience with AI text generation models, past programming knowledge (Likert scores 1-5; 5 highest), and whether they had ever worked on a project involving evaluating LLMs. Participants then watched a five-minute video introducing the interface. 

In the structured task, participants navigated a mock prompt engineering scenario in two parts, where a developer first chooses a model, then iterates on a prompt to improve performance according to some criteria. We asked participants to choose a model and prompt to ``professionalize an email'' (translate a prospective email message to sound more professional).\footnote{Although we could have contrived a task with an objective `best' answer---best model or prompt template---that wouldn't reflect the kind of ambiguities present in many real world decisions around LLM usage.} In part one, participants were given a preloaded flow, briefed on the scenario (\quo{Imagine you are a developer...}), and presented with two criteria on a slip of paper: (1) \emph{The response should just be the translated email}, and (2) \emph{The email should sound very professional}. Participants were tasked with choosing the `best' model given the criteria, and to justify their choice. All participants saw the exact same cached responses from GPT-4, Claude-2, and PaLM2, in the exact same order, for the prompt \quo{Convert the following email to have a more professional and polite tone} with four example emails (e.g., \quo{Why didn't you reply to my last email???}). After they spent some time inspecting responses, we asked them to add one more example to translate and to increase \emph{Num of responses per prompt}, to show them how the same LLMs can vary on the same prompt. 

Once participants chose a model, we asked them to remove all but their selected model. We then guided them to abstract the pre-given ``command prompt'' into a TextFields, and add at least two more command prompts of their own choosing. On a slip, we gave them a third criteria: \quo{the email should be concise.} %
After participants inspected responses and started to decide on a `best' prompt, we asked them to add one code Evaluator and Vis Node, plotting lengths of responses by their \emph{command} variable. After spending some time with the plot, participants were asked to decide.

The remaining study time was taken up by an unstructured, exploratory section meant to emulate how users---provided enough support and documentation---might use \oursystem{} to investigate a hypothesis about LLM behavior that mattered to them. We asked participants a day before their study to think up an idea, question, or hypothesis they had about AI text generation models, and gave a list of six possible investigation areas (e.g., checking models for bias,  conducting adversarial attacks), but did not provide any concrete examples. During the study, participants then explored their idea through the interface with the help of the researcher. Importantly, researchers were instructed to only support participants in pursuit of their investigations, not to guide them towards particular domains of interest. %
The one exception is where a participant only queried a single model; in this case, the researcher could suggest that the user try querying multiple models at once. Participants used the exact same interface as the public version of our tool, and had access to OpenAI's \emph{gpt-3.5} and \emph{gpt-4}, Anthropic's \emph{claude-2}, Google's \emph{chat-bison-001}, and HuggingFace models. 

After the tasks, we held a brief post-interview (5-10 min), asking participants to rate the interface (1-5) and explain their reasoning, what difficulties they encountered, suggestions for improvements, whether they felt their understanding of AI was affected or not, and whether they would use the interface again and why.

\subsection{Recruitment, Participant Demographics, and Data Analysis}

We recruited in-lab participants around our U.S.-based university through listservs, Slack channels, and flyers. We tried to expand our reach beyond people experienced in CS and ML, specifically targeting participants in humanities and education. Participants were generally in their twenties to early thirties (nine 23-27; eight 28-34; three 18-22; one 55-64), predominantly self-reported as male (14 men, 7 women), and largely had backgrounds in computing, engineering, or natural sciences (ten from CS, data science, or tech; seven from bioengineering, physics, material science, or robotics; two from education; one from medicine and one from design). They had a moderate amount of past experience with AI text generation models (mean{=}3.3, stdev{=}1.0); one had none. Past Python programming experience varied (mean{=}3.1, stdev{=}1.3), with less experience in JavaScript (mean{=}2.0, stdev{=}1.3); two had no programming experience. Eight had ``worked on an academic study, paper, or project that involved evaluating large language models.'' All participants came in to the lab, with studies divided equally among the first three coauthors. Each study took 75 minutes, and participants were given \$30 in compensation (USD). Due to ethical concerns surrounding the overuse of Amazon gift cards in human subject studies \cite{pater2021standardizing, ngbeyond}, we paid all participants in cash. 

For our interview study, we sought participants who had already used \oursystem{} for real-world tasks, reaching out via social media, GitHub, and academic networks. The first author held six semi-structured, 60 min. interviews with eight participants (in two interviews, two people had worked together). Interviews took place via videoconferencing. Interviewees were asked to share their screen and walk through something they had created with \oursystem{}. Unlike our in-lab study, we kept interviewees' screen recordings private unless they allowed us to take a screenshot, since real-world users are often working with sensitive information. Interviewees generously volunteered their time. 

We transcribed all 32 hours of screen recordings and interviews, adding notes to clarify participant actions and references (e.g.,  \quo{[Opens inspector; scrolls to top]. It seems like it went fast enough... [Reading from first email group] `Hi...'}). %
We noted conceptual or usability problems and the content of participant references. %
We analyzed the transcripts through a combination of inductive thematic analysis through affinity diagramming, augmented with a spreadsheet to list participants' ideas, behaviors (nodes added, process of their exploration, whether they imported data, etc), and answers to post-interview questions. For our in-lab study, three coauthors separately affinity diagrammed three transcripts each, then met and joined the clusters through mutual discussion. The merged cluster was iteratively expanded with more participant data until clusters reached saturation. For interviews, the first author affinity diagrammed all transcripts to determine themes. In what follows, in-lab participants are P1, P2, etc.; interviewees are Q1, Q2, etc.

\section{Modes of Prompt Engineering and LLM Hypothesis Testing} \label{modes}

What process do people follow when prompt engineering and testing hypotheses about LLM behavior more generally? Before we break down findings per study, we provide a birds-eye view of how participants in general used \oursystem{}. Synthesizing across studies, we find that people tend to move from an \emph{opportunistic exploration} mode, to a \emph{limited evaluation} mode, to an \emph{iterative refinement} mode. About half of our in-lab users, especially end-users with limited prior experience, never left exploration mode; while programmers or auditors of LLMs quickly moved into limited evaluation mode. Some interviewees had disconnected parts of their flows that corresponded to exploration mode, then would scroll down to reveal extensive evaluation pipeline(s), explaining they had transferred prompts from the exploratory part into their evaluation. In Appendix~\ref{case_studies}, we provide one Case Study for each mode. Notice how these modes correspond to users moving from the left side of Fig.~\ref{fig:llm-ops} towards the right. 

{\bf Opportunistic exploration mode} is characterized by rapid iteration on prompts, input data, and hypotheses; a limited number of prompts and input data; and multi-model comparison. Users {\bf \emph{prompt / inspect / revise}}: send off a few prompts, inspect results, revise prompts, inputs, hypotheses, and ideas. In this mode, users are sending off quick experiments to probe and poke at model behavior (\quo{throw things on the wall to see what's gonna stick}, Q3). For instance, participants who conducted adversarial attacks like jailbreaking \cite{deng2023jailbreaker} %
would opportunistically try different styles of jailbreak prompts, and were especially interested in checking which model(s) they could bypass.

{\bf Limited evaluation mode} is characterized by moving from ad-hoc prompting to \emph{prototyping an evaluation}. Users have reached the limits of manual inspection and now want a more efficient, ``at-a-glance'' test of LLM behavior, achieved by encoding criteria into automated evaluator(s) to score responses. Users {\bf \emph{prompt / evaluate / visualize / revise}}: prompt model(s), score responses downstream in their chain, visualize results, and revise their prompts, input data, models, and/or hypotheses accordingly. Hallmarks of this mode are users setting up an analysis pipeline, iterating on their evaluation itself, and ``scaling up'' input data. The evaluation is ``limited'' as evaluations at this stage are often ``coarse''---for example, rather than checking factuality, check if the output is formatted correctly.\footnote{Crucially, this mode does \emph{not} imply multiple prompts: a few in-lab participants set up an evaluation pipeline that \emph{only sent off a single prompt}. %
Though these participants did add a TextFields or Tabular Data node, it only had a single value/field. Indications were that, with more time, they would have ``scaled up'' how many inputs or parameters they were sending to test more specific hypotheses. However, some users might have also had conceptual trouble imagining how to scale up their testing; we discuss this more later.} 

{\bf Iterative refinement mode} is characterized by having an already-established evaluation pipeline and criteria and \emph{tweaking} prompt templates and input data through further parametrization or direct edits, setting up one-off evaluations to check effects of tweaks, increasing input data complexity, %
and removing or swapping out models. Users {\bf \emph{tweak / test / refine}}: modify or parametrize some aspect of their pipeline, test how tweaks affect outputs compared to their ``control'', and refine the pipeline accordingly. The key difference between limited evaluation and iterative refinement is in the solidity of the chain: here, users' prompts, input data, and evaluation criteria have largely stabilized, and they are looking to \emph{optimize} (e.g., through tweaks to their prompt, or extending input data to identify failure modes). Some interview participants had reached this mode, and were refining prompt templates or scaling up input data. The few in-lab participants that had brought in ``prompt engineering'' problems by importing prompt templates or spreadsheets would immediately set up evaluation pipelines, moving towards this mode.

\vspace{0.06in}

These modes are suggestive and not rigidly linear; e.g., users may scrap their limited evaluation and return to opportunistic exploration. In Sections~7 and~8 below, we delve into specific findings for each study. For our in-lab study, we
describe how people selected prompts and models, how \oursystem{} supports exploration and understanding, and note conceptual and usability issues. For our interview study, we focus on what differed from in-lab users. %

\section{In-lab Study Findings}

On average, participants rated the interface a 4.19/5.0 (stdev{=}0.66). No participant rated it lower than a three. When asked for a reason for their score, participants generally cited minor usability issues (e.g., finicky when connecting nodes, color palette, font choice, more plotting options). Eighteen participants wanted to use the interface again; five before being explicitly asked. %
Some just wanted to play around, %
citing model comparison and multi-response generation. %
Participants who had prior experience testing LLM behavior in academia or industry cited speed and efficiency of iteration as the primary value of the tool (\quo{If I had started with using this, I'd have gotten much further with my prompt engineering... This is much faster than a Jupyter Notebook}, P4; \quo{this would save me half a day for sure... You could do a lot of stuff with it%
}, P21). Participants mentioned prior behavior as having multiple tabs open to chat with different models, manually copying responses into spreadsheets, or writing programs. %
Three wanted to use \oursystem{} for research. 

We recount participants' behavior in the structured task to choose a model and prompt template, overview how \oursystem{} supported participants' explorations and understanding, and reflect on pain points.

\subsection{How People Decide on Models and Prompts}

How do people choose a text generation model or prompt, when presented with side-by-side responses? People appear to weigh \emph{trade-offs} in response quality for different criteria and contexts of usage. Participants would perceive one prompt or model to excel in one criteria or context, but do poorly in another; for another prompt or model, it was vice-versa. Here, we use ``criteria'' liberally to mean both our explicit criteria and also participants' tacit preferences. %
Participants would also implicitly \emph{rank} criteria, assigning more weight to some over others, and refer to friction between criterias (e.g., P2 \quo{prefer[red] professional over concise, because it [email] can be concise, but misconstrued}). Moreover, seeing \emph{multiple representations} of prompt performance, each of which better surfaced aspects of responses that corresponded to different criteria, could affect participants' theorizing and decision-making. We unpack these findings here.

For the first part of our structured task, participants reached no consensus on which model performed ``better'': eight chose PaLM2, seven GPT-4, and six Claude-2. There was no pattern in reasoning. Participants \emph{did} notice similar features of each models' response style, but \emph{how} they valued that style differed. Some participants liked some models for the same reason others disliked them; for instance, P1 praised PaLM2 for its lengthy emails; while P17 chose GPT-4 because \quo{PaLM2 is too lengthy.} Although we had deliberately designed our first criteria against the outputs of Claude (for its explanatory information around the email), some participants still preferred Claude, perceived its explanations as useful to their imagined users, or preferring its writing style. In the unstructured task, participants developing apps also mentioned exogenous factors such as pricing, access, and response time when comparing models.

\begin{figure}
  \fbox{\includegraphics[width=\columnwidth]{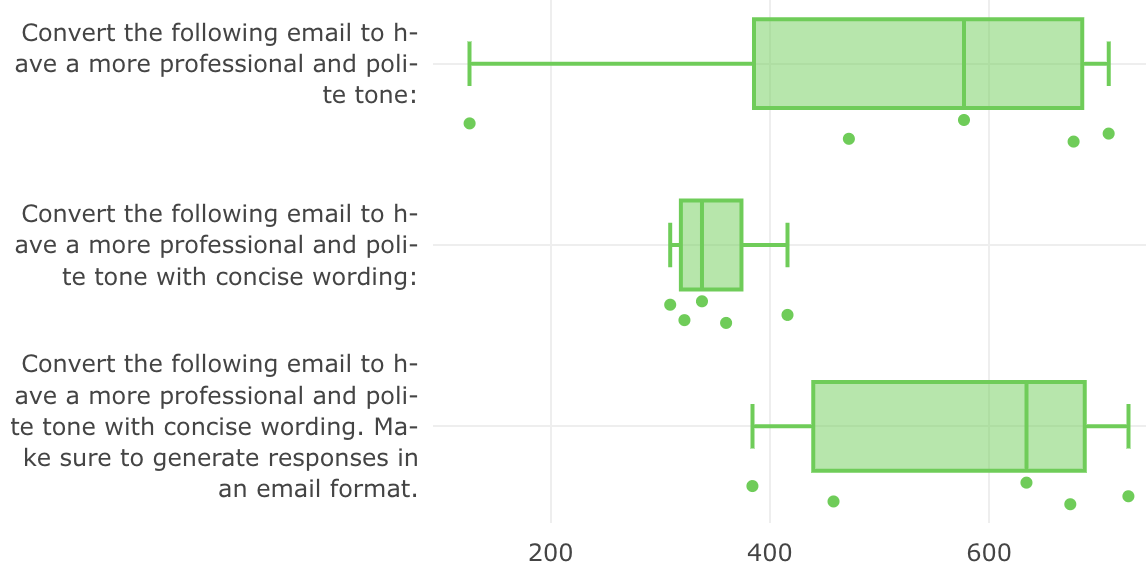}}
  \caption{P17 plots the response lengths of three command prompts, augmenting her theories about each prompts' performance.}
  \Description{}
  \label{fig:plotting-prompts}
\end{figure}

How did people choose one prompt among multiple? Like when choosing models, participants appeared to weigh trade-offs %
between different criteria and contexts. %
Having multiple representations (e.g., plots of prompt performance) could especially give users a different ``view'' that augmented understanding and theorizing. P1 describes tensions between his and his users' needs, referencing both manual inspection and a plot of response lengths by prompt:

\begin{quote}
    \quo{If I am a developer, I like this one [third prompt] because it will help me better to pass the output... But if they [users] have a chance to see this graph [Vis node], they would probably choose this one [second prompt] because it fits their needs and it's more concise [box-and-whiskers plot has smallest median and lowest variability]... So I think it depends on the view.}
\end{quote}

Multiple representations could also augment users' theorizing about prompting strategy. For instance, P17 had three command prompts, each iteration just tacking more formatting instructions onto the end of the default prompt (Figure~\ref{fig:plotting-prompts}). Comparing between her plot and Table Layout, she theorizes: \quo{After adding `generate response in an email format' it made it lengthier... %
But if I don't say `with concise wording'... %
sometimes it generates responses that are %
three paragraphs, for a really simple request. So I would [go with] the second instruction... %
[and its] the length difference [variance] is less.} Seeing that one prompt resulted in shorter or \emph{less variable} responses could cause a participant to revise an earlier opinion. After noticing via the plot that his first command \quo{seem[s] more consistent}, P4 wanted to mix features from it into his chosen prompt to improve the latter's concision, as he still preferred the latter's textual quality.

These observations suggest that systematic evaluations can contest fixation \cite{zamfirescu2023whyjohnny} caused by manual inspection. However, it also reveals users may need \emph{multiple} representations, or they will make decisions biased by features that are easiest to spot in only one. With multiple, they can make decisions more confidently, mixing and matching parts of each prompt to progress towards an imagined ideal. The benefit of prompt comparison also underscores the importance of starting from a \emph{variety} of prompts---similar to past work \cite{zamfirescu2023whyjohnny}, many of our participants struggled to come up with a variety of prompts, with thirteen just perturbing our initial command prompt. We reflect on this more in our Discussion. 

\subsection{\oursystem{} Supported a Variety of Use Cases and Users}

Participants brought in a variety of ideas to the unstructured task, ranging from auditing of LLM behavior to refining an established prompt used in production. We recount three participants' experiences as Case Studies in Appendix~\ref{case_studies}, each corresponding to a Mode of Usage from Section~\ref{modes}. Seven participants evaluated model behavior given concrete criteria, with six importing prior data; ideas ranged from testing model's ability to understand program patch files, to classifying user attitudes in messaging logs. Nine audited models in opportunistic exploration mode, looking for biases or testing limits (e.g., asking undecidable questions like \quo{Does God exist?}, P20). Of these users, four conducted adversarial attacks \cite{deng2023jailbreaker}, seemingly influenced by popular culture about jailbreaking. P9 and P15, both with no programming experience, used the tool to audit behavior, the former comparing models' ability to generate culturally-appropriate stories about Native Alaskans.  %
Others were interested in generating text for creative writing tasks like travel itineraries. %
Participants often searched the internet, such as cross-checking factual data, copying prompts from Reddit, or evaluating code in an online interpreter. Overall, nine participants imported data (six with spreadsheets) to use in their flow. %

\subsection{\oursystem{} Affected Participants' Understanding of AI Behavior or Practice}

In the post-interview, fifteen participants said their understanding of AI was affected by their experience. Six were surprised by the performance of Claude-2 or PaLM2, feeling that, when confronted with direct comparisons to OpenAI models, they matched or exceeded the latter's performance. %
Five said that their \emph{strategy} of prompting or prompt engineering had changed (\quo{[Before], I wasn't doing these things efficiently... I [would] make minor modifications and rerun, and that would take hours... %
Here, since everything is laid out for me, I don't want to give up}, %
P4). Others less experienced with AI models learned about general behavior. P16, who had never prompted an AI model before, \quo{realized that different models have completely different ways of understanding my prompts and hence responding, they also have a completely different style of response.} P15, covered in Case Study~\ref{case1}, said she had lost \quo{trust} in AI. %

\subsection{Challenges and Pain-points}

Though many participants derived value from \oursystem{}, that is not to say their experience was frictionless. %
The majority of usability issues revolved around the flow UI, such as needing to move nodes around to make space, connecting nodes and deleting edges; others %
related to inconsistencies in the ordering of plotted variables, and wanting more control over colors and visualizations. %
Some participants also encountered conceptual issues, which sometimes indicate users getting used to the interface. The most common conceptual issue was learning how prompt templating worked, and especially, forgetting to declare input variables in Prompt Nodes. %
Once users learned how to template, however, the issue often disappeared (\quo{prompt variables... there's a bit of a learning curve, but I think it makes sense, the design choice}, P13). Learning template variables seemed related to past programming expertise and not AI, suggesting users without any prior programming experience will need extra resources.\footnote{For instance, P16, who had never prompted an AI model before, attributed her adeptness with templating to prior programming experience. By contrast, P9 had no programming experience and struggled; after the intro video he was \quo{a little overwhelmed}, asking: \quo{what language am I in?}} 

Import to reflect on is that, in the lab, researchers were on-hand to guide users. Although users were the ones suggesting ideas---often highly domain-specific ones---researchers could help users with ways to implement them and overcome conceptual hurdles. Some end-users \emph{and even a few users with substantial prior experience with AI models or programming with LLM APIs} appeared to have trouble ``scaling up,'' or systematizing, their evaluations. For example, P10 rated themselves as an expert in Python (5) and had conducted prior research on LLM image models. They set up an impressive evaluation, complete with a prompt template, Prompt Node, Chat Turn, Simple Evaluator and Vis nodes, but ultimately only sent off a single prompt to multiple models. We remark more on this behavior in Discussion.

\section{Interviews with Real-World Users}

Our interview findings complement, but in important places diverge, from our in-lab studies. Like many in-lab participants, real-world users praised \oursystem{}'s features and used it for goals we had designed for---like selecting models or prompt testing---however, some things real users cared about were hardly, if ever mentioned by in-lab participants. As we analyzed the data and compared it with our in-lab study, we realized that many user needs and pain-points revolve around the fact that they were using \oursystem{} to \emph{prototype data processing pipelines}, a wider context that re-frames the tasks we had designed \oursystem{} to support as subtasks of a larger goal. 
Interviewees remarked most about \emph{easing the export and sharing of data} from \oursystem{}, %
adding \emph{processor nodes}, the \emph{importance of the Inspect Node} for sharing and rapid iteration, and the \emph{open-source nature} of the project for their ability to adapt the code to their use case. We discuss these insights more below; but first, we provide an overall picture, reviewing similarities, use cases, and concrete value that real-world users derived from \oursystem{}. %

\begin{table*}[t]
    \centering
    \small
    \begin{tabularx}{\textwidth}{|l|p{0.36in}|p{0.3in}|p{0.5in}|p{1.0in}|X|X|X|}
    \hline
    ID(s) & Location & Work & How discovered & Use Case(s) & Major nodes and features used & Outcome \\
    \hline
    Q1-2 & U.S. (west) & Acad. & Medium post & Adapting code for LLM image model prototyping & \emph{Adapted source code (esp. nodes, model querying, cacheing)} & Submitting HCI paper to major conference in time for deadline \\
    \hline
    Q3 & U.K. & Ind. & GitHub & Supporting requirements analysis in software testing & Prompt, Tabular Data,  TextFields, Vis, JS Eval; export to excel, metavariables & \quo{I've had good conversations with [other] developers as a result} \\
    \hline
    Q4 & U.S. (east) & Acad. & Hacker-News & Auditing models for gender bias & Prompt, TextFields, Chat Turn, Inspect node; prompt chaining, LLM scoring & Improved prompt templates for a pipeline they are building in Python \\
    \hline
    Q5-6 & Germany & Ind. & LearnPro-mpting.org & Information synthesis and extraction from documents & Prompt, Tabular Data, JS Eval, Vis,  Comment, Inspect node; comparing models & \quo{Convinced client [to] continue with the next phase of [a] project} \\
    \hline
    Q7 & U.S., Austria & Acad. & Word of mouth & Cleaning a column of tabular data & Prompt, Tabular Data, TextFields, JS Eval, Vis; variables in code eval & Decided that LLMs were not reliable enough  for their task \\
    \hline
    Q8 & U.S. (central) & Acad. & Twitter & Extracting and formatting info from podcast episode metadata & Prompt, TextFields, CSV, JS Eval, Vis, Comment; comparing models, prompt chaining, template chaining & Found that neither OpenAI model produced good enough outputs; now looking into  local models \\
    \hline
    \end{tabularx}
    \caption{Our interviewees. All interviewees imported data into \oursystem{}, with the exception of Q1-2 (who adapted source code). Interviewees commonly had multiple evaluation branches in their flows, as well as multiple flows in the same document.}
    \label{tab:interviewees}
\end{table*}

We list interview participants in Table~\ref{tab:interviewees}, with use cases and nodes used. The Outcome column suggests the actionable value that \oursystem{} provided. Note that Q1 and Q2's primary use case was building on the source code to enable their HCI research project. All six users of the \emph{interface} found it especially useful for prototyping and iterating on prompts and pipelines (e.g., Q5: \quo{I see the use case for \oursystem{} as a very good prompt prototyping environment}). Usage reflected modes of \emph{limited evaluation} and \emph{iterative refinement}, with multiple participants describing a prompt/evaluate/visualize/revise loop: query LLM(s), evaluate responses and view the plot, then refine prompts or change models, until one reaches the desired results. For instance, Q3 described tweaking a prompt template until the LLM output in a consistent format, facilitated by maximizing 100\% bars in a Vis Node across all input data. Some participants saw \oursystem{} as a rapid prototyping tool missing from the wider LLMOps ecosystem, a tool they used \quo{until I get to the point where I can actually write it into hard code} (Q4). Three appreciated how \emph{few} nodes there were in \oursystem{} given its relative power, compared to other node-based interfaces (e.g., Q8: \quo{It's impressive. What you're able to accomplish with so few}). 
They worried that adding too many new nodes would make the interface more daunting for new users. Q4 and Q7 found it more effective than Jupyter notebooks (Q7: \quo{I enjoyed \oursystem{}... because I could run the whole workflow over and over again, and... in Jupyter, that was not easy}). In the rest of this section, we expand upon differences from our in-lab study. 

\subsection{Prototyping data processing pipelines}

Five interviewees were using \oursystem{} not (only) for prompt engineering or model selection, but for \emph{on-demand prototyping of data processing pipelines involving LLMs}. All imported data from spreadsheets, then would send off many parametrized prompts, iterate on their prompt templates and pipelines, and ultimately export data to share with others. Such users also used \oursystem{} for at least one of its intended design goals, but always in service of their larger data processing goal.
For Q7, \quo{the idea was to write a pipeline that... helps you with this whole process of data cleaning.} For him, \oursystem{} was ideal for \quo{whenever you have a variety of prompts you want to use on something particular, like a data set. And you want to explore or investigate something.} Another user, Q3, would open his refined flow, edit one value, re-run it and then export the responses to a spreadsheet. Like other participants, he remarked on \oursystem{}'s combinatorial power as its chief benefit, compared to other tools (\quo{This tool is strong at prompt refining. With [Flowise]...Let's say I wanted to try multiple [input fields]. I don't think I could do that}). Participants also mentioned iterating on the \emph{input data} as part of the prototyping process. Finally, related to data processing, three users wished for processor nodes, like ``join'' nodes to concatenate LLM responses, and in one case were manually copying LLM outputs into a separate flow to emulate concatenation. Note that many needs and pain-points below are related to data processing. 

\subsection{Getting data out and sharing with others}

Many participants wanted to export data out of \oursystem{}. This was also the most common pain point, especially when transitioning from a prototyping stage---which they perceived as \oursystem{}'s strong point---to a production stage (e.g., \quo{it would be helpful when we are out of this prototyping stage, that the burden or the gap---changing the environment... gets tightened}, Q5).
Needs broke down into two categories: exporting for integration into another application, and exporting for sharing results with others. For the former, developer users would use \oursystem{} to battle-test prompts, model behavior, and/or prompt chains, but then wished for an easier way to export their flows to text files or app building environments.\footnote{This does not, however, mean they perceived \oursystem{} as an ``app building'' tool---%
some even expressed worry about it trying to do too much. When asked about how he would feel if \oursystem{} supported app development, Q3 remarked, \quo{I just worry about it [\oursystem{}] becoming too complicated. Like, are you building the app side of it?''} He said even if \oursystem{} supported app-building, it would need different ``modes'': \quo{app building mode or prompt refining mode.}} For the latter, five interviewees shared results with others, whether through files, screenshots of their flows, exported Excel spreadsheets of responses, or copied responses. Q5 and Q6 stressed the importance of the Inspect Node---a node that no in-lab participant used or mentioned (\quo{[Once] the result is worth documenting, you create an Inspect node.}). They took screenshots of flows and sent them to clients, in one case convincing a client to move forward with a project. The anticipation of sharing with others also could change behavior. Q3 had several TextFields nodes with only a single value, \quo{because I knew that it was something that essentially other teams might want to change.}. %
Sharing could also be a pain-point, with two wanting easier shareable ``reports'' of their analysis results.

\subsection{Pain points: Hidden affordances and friction during opportunistic exploration mode} \label{pain}

Like in-lab participants, interviewees also encountered usability and conceptual issues. A common theme was individual users expressing a need for features that already exist but are relatively hidden, surfaced only through examples or documentation. These hidden affordances included \emph{implicit template variables}, \emph{metavariables}, and \emph{template chaining}. The former two features address users' need to reference upstream metadata---metadata associated with input data or responses---further downstream in a chain.\footnote{For example, in \oursystem{} one can define a column of a table and then refer to it downstream via a metavariable (Fig.~\ref{fig:tabular-data}). However, Q5 and Q6 seemed unaware of this feature and had implemented a workaround. For implicit template variables, Q4 needed to reference an upstream value \{gender\} later downstream in a prompt chain, and was unaware that an implicit template variable could accomplish this (e.g. \{\#gender\}).} Another pain point was friction during the opportunistic exploration phase. In Section~6, we mentioned some interviewees had disconnected regions of their flows, with one region we termed opportunistic exploration mode (rapid, early-stage iteration through input data, prompts, models, and hypotheses; usually, a chain of three nodes, TextField-Prompt-Inspect). In this mode, some interviewees preferred to inspect responses directly on the flow with an Inspect Node (instead of the pop-up window), as it facilitated rapid iteration. They wanted an even more immediate, in-context way to read LLM responses that would not require them to attach another node.\footnote{This need was again reflected in a later GitHub Issue and has since been addressed with a pull-out inspector drawer.}

\subsection{Open-source flexibility}

Multiple interviewees mentioned looking at our source code, and two projects extended it. Q5 and Q6, employees of a consulting firm that works with the German government, extended the code to support a German-based LLM provider, AlephAlpha \cite{aleph-alpha}, complete with a settings screen. They cited the value of supporting European businesses and GDPR data protection laws: \quo{the government [of Germany] wants to support it. It's a local player... [and] There's a strong need to to hide and to to protect your data. I mean, GDPR, it's very strict in this.} Their goal was to use \oursystem{} to  determine ``if it makes sense to switch to'' the German model for their use cases, over OpenAI models. HCI researchers Q1 and Q2's chief interaction with the tool was its source code, finding it helpful for jumpstarting a project on a flow-based tool for LLM image model prototyping. Q2 appreciated the \quo{thought put into} caching, Prompt Node progress bar, and multi-model querying, adding: \quo{It was very easy for me to set up \oursystem{}... [and it was] surprisingly easy to [extend]... a lot easier than I had expected.} They said that the jump-start \oursystem{} provided was a chief reason they were able to complete their project in time to submit a paper to the annual CHI conference. 

\section{Discussion and Conclusion}

Our observations suggest that \oursystem{} is useful both in itself, but also as an `enabling' contribution, an open-source project which others can extend (and are extending) to investigate their own ideas and topics, including other research publications to this very conference. Given that \oursystem{} was released only a few months ago, we believe the stories presented here provide evidence for its real-world usefulness. In what follows, we review our key findings.

Our work represents one of the only ``prompt engineering'' system contributions with data about real-world usage, as opposed to in-lab studies on structured tasks. Some of what real users cared about, like features for exporting data and sharing, were absent from our in-lab study---and are, in fact, also absent from similar LLM-prompting-system research with in-lab studies \cite{wu2022promptchainer, wu2022ai, brade2023promptify, mishra2023promptaid, zamfirescu2023whyjohnny}. %
Most surprising (to us) was that some knowledge workers were using \oursystem{} for a task we had never anticipated---\emph{data processing}. Although we only had six interface users in our interview study, %
the only two in-lab participants in startups, P8 and P4, were both testing LLMs' ability to process and reformat data.
Most prior LLM tools target sensemaking \cite{jiang2023graphologue, suh2023sensecape}, prompt engineering \cite{mishra2023promptaid, jiang2022promptmaker}, or app building \cite{wu2022promptchainer}, but do not specifically target, or even mention, data processing. Our findings suggest a need for %
systems to support \emph{on-demand creation of data processing pipelines involving LLMs,} where the purpose is not (always) to make apps, but simply process data and share the results. %
\oursystem{}'s combinatorial power---the ability to send off many queries at once, parametrized by imported data%
---appeared key to supporting this need. Future systems should go further by providing users more accessible ways to reference upstream metadata further downstream in their chain (see \ref{pain}).

Second, we identified three modes of prompt engineering and LLM hypothesis testing: \emph{opportunistic exploration}, \emph{limited evaluation}, and \emph{iterative refinement}. The first mode is similar to Barke et al.'s exploration mode for GitHub CoPilot \cite{barke2023grounded}. %
Future systems should explicitly consider these modes when designing and framing the work. For instance, users often too quickly enter iterative refinement mode---refining on the first prompt they try---rather than exploring a variety before settling on one \cite{zamfirescu2023whyjohnny}. If a prompt engineering tool only targets iterative refinement, then the opportunistic exploration stage---finding a good prompt to begin with---may be too quickly skirted over, trapping users in potentially suboptimal prompting strategies. %
These modes also suggest design opportunities. For instance, we believe that \oursystem{}'s design could have better supported opportunistic exploration mode, with some users wanting a simpler way to inspect LLM responses in-context (\ref{pain}). One design solution may be to concretize each mode into separate, related interfaces or layouts---e.g., a more chat-like interface for exploration mode, that then facilitates the transition to later modes, each with dedicated interfaces. Prior LLM-prompting systems seem to target opportunistic exploration \cite{jiang2023graphologue, suh2023sensecape} or iterative refinement \cite{mishra2023promptaid, strobelt2022interactive}, but overlook \emph{limited evaluation}: an important mid-way point characterized by prototyping small-scale, quick-and-messy evaluations on the way to greater understanding. Future work might target the prototyping of on-demand LLM evaluation pipelines themselves (see ``model sketching'' for inspiration \cite{lam2023model}). 

Third, we found that when people choose different prompts and models, they weigh \emph{trade-offs} in performance for different criteria and contexts, and bring their own perspectives, values, preferences, and contexts to bear on decision-making. Having multiple representations of responses seemed to help participants weigh trade-offs, rank prompts and models, develop better mental models, and make revisions to their prompts or hypotheses more confidently. Connecting to theories of human learning~\cite{gentner1997structure, marton2014necessary}, the case study in \ref{case_studies}.1 suggests that cross-model comparison might also help novices improve mental models of AI by forcing them to encounter differences in factual information, jarring AI over-reliance~\cite{liao2022designing}. The subjectivity of choosing a model and prompt implies that, while LLMs can certainly \emph{help} users generate or evaluate prompts \cite{brade2023promptify, zhou2022large}, there will never be such a thing as \emph{fully} automated prompt engineering. Rather than framing prompt engineering (purely) as an optimization problem, projects looking to support prompt engineering should instead look for ways to give users greater \emph{control} over their search process (e.g., ``steering'' \cite{brade2023promptify, zhou2022large}). 

A final point and caveat: while users found \oursystem{} useful for \emph{implementation} and \emph{iteration}, including on real-world tasks, more work needs to be done on \emph{conceptualization} and \emph{planning} aspects, to help users move out of opportunistic exploration into more systematic evaluations. In-lab users seemed limited in their ability to imagine systematizing their tests, \emph{even a few with prior expertise in AI or programming with LLM APIs}. %
This extends prior work studying how ``non-AI-experts'' prompt LLMs~\cite{zamfirescu2023whyjohnny}, suggesting even people who otherwise perceive themselves to be AI experts may have trouble systematizing their evaluations. Since LLMs are nondeterministic (at least, often queried at non-zero temperatures) and prone to unexpected jumps in behavior from small perturbations, it is important that future systems and resources help reduce fixation and guide users from early exploration into systematic evaluations. We might leverage concepts from tools designed for more targeted use cases; e.g., the auditing tool AdaTest++ provides users ``prompt templates that translate experts' auditing strategies into reusable prompts'' \cite[p. 15-6]{rastogi2023supporting}. Other work supports creation of prompts or searching of a ``prompt space'' \cite{shi2023prompt, mishra2023promptaid, strobelt2022interactive}. To support systematization/scaling up, we might also employ an interaction whereby a user chats with an AI that sketches out an evaluation strategy.

\subsection{Limitations}

Our choice to use a qualitative evaluation methodology derived from well-known difficulties around toolkit research \cite{ledo2018evaluation, olsen2007evaluating}, concerns about ecological validity, and, most importantly, from the fact that we could not find a prior, well-established interface that matched the entire featureset of \oursystem{}. Our goal was thus to establish a baseline system that future work might improve upon. While we believe our qualitative evaluation yielded some important findings, more quantitative, controlled approaches should be performed on parts of the \oursystem{} interface to answer targeted scientific questions. Our in-lab study was also of a relatively short duration (75 min); future work might observe changes in user behavior over longer timeframes, for instance with a multi-week workshop. Finally, for our interview study, we acknowledge a self-selection bias, where participating interviewees may already have found \oursystem{} useful, missing users who did not. Our in-lab study provided some insights---we speculate that users' prior exposure to programming was important to the quality of their experience.  

\begin{acks}
This work was partially funded by the NSF grants IIS-2107391, IIS-2040880, and IIS-1955699. Any opinions, findings, and conclusions or recommendations expressed in this material are those of the author(s) and do not necessarily reflect the views of the National Science Foundation.
\end{acks}

\bibliographystyle{ACM-Reference-Format}
\bibliography{citations}

\appendix

\section{Case Studies for Modes of Usage} \label{case_studies}

To help readers understand how people used \oursystem{} and how their interactions varied, we walk through three participants' experiences. Each Case Study illustrates one mode from Section~\ref{modes}. %

\begin{figure*}
  \centering
  \includegraphics[width=\textwidth]{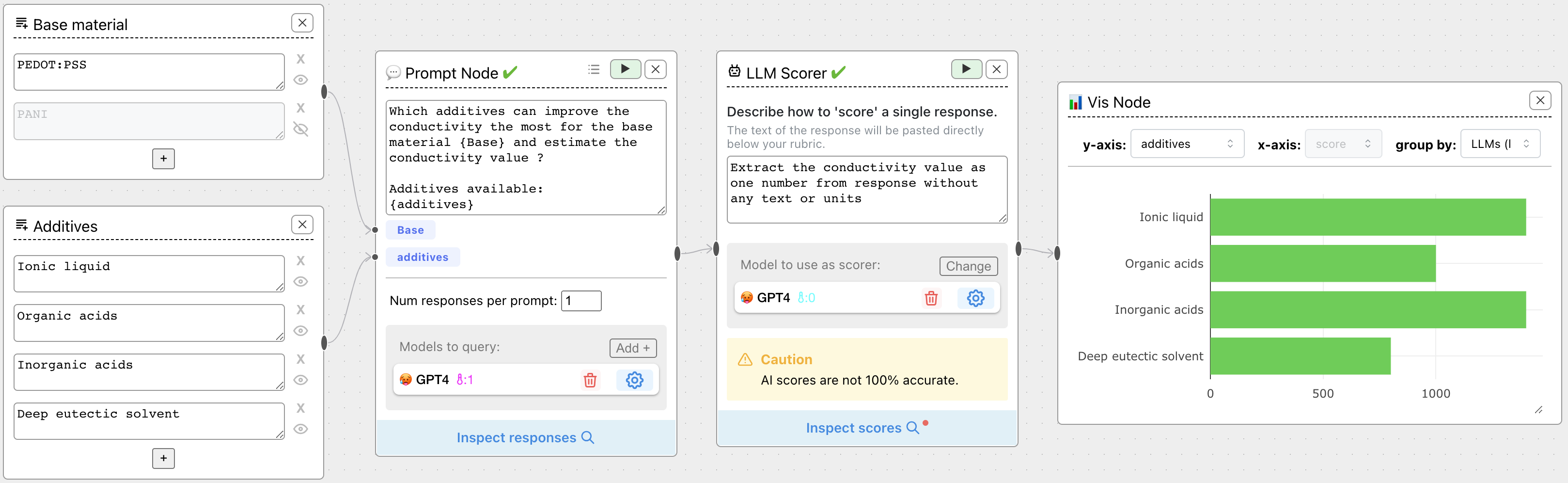}
  \caption{P18's final flow, with one value toggled off to display their initial plot. The user asked GPT-4 to estimate how much the conductivity to a base polymer, \texttt{PEDOT:PSS}, may increase given one of four additives. They use an LLM Scorer to extract the mentioned conductivity value; after spot-checking it using Inspect, they plot in a Vis Node, finding it \quo{roughly} correct.} %
  \Description{}
  \label{fig:limited-eval}
\end{figure*}

\subsection{Opportunistic exploration mode: Iterating on hypotheses through rapid discovery of model behavior.} \label{case1}
A graduate student from Indonesia, P15 wanted to test how well AI models knew the Indonesian education participation rate and could give advice on %
\quo{what the future of us as educators need to do.} She opens a browser tab with official data from Badan Pusat Statistik (BDS), Indonesia's Central Agency for Statistics. %
She wants to know \quo{what is the difference, if I use a different language?} %
She adds a TextFields with two fields, one prompt in English, \quo{Tell me the participation rate of Indonesian students going to university}; 
the second its Indonesian translation. %
\quo{Let's just try two. I just want to see where it goes.} Collecting responses, she looks over side-by-side responses of three models to her English prompt. All models provide different years and percentages. %
Scrolling down and expanding the response group for her Indonesian prompt, she finds that Falcon.7B only repeats her prompt and the PaLM2 model has triggered a safety filter.\footnote{This is a real problem with PaLM2 that we have communicated with the Google AI team about; they identified the issue and are fixing it.} The last model, GPT-3.5, gives a different statistic than its English response. %

Looking over these responses in less than a minute, P9 has discovered three aspects of AI model behavior: first, that models differ in their ``facts''; second, that some models can refuse to answer when queried in a non-English language; third, that the \emph{same} models can differ in facts when queried in a different language. She compares each number to the BDS statistics, finding them inaccurate. \quo{Oh my god, I'm curious. Why do they have like different answers across [models]?} She then adds models to the Prompt Node. \quo{Can I try all [models]? I %
want to see %
if it's in the table.} 

She queries the new models. A new hypothesis brews: \quo{In our prompt, [do] we need to say our source of data? Would that be like, more accurate?} She wonders if different models are pulling data from different sources. Inspecting responses, she finds some models have cited sources of data: Claude cites UNESCO and GPT-4 cites the World Bank, UNESCO, and %
the Indonesian Ministry of Education and Culture. %
For her Indonesian prompt, she discovers that the same models only cite BPS in their responses. \quo{BPS is only mentioned when I use Indonesian... %
For the English [prompt]... %
[it's] more like, global... Wow, it's very interesting how, the different language you use, there's also a different source of data.}

She adds a second prompt variable, \emph{\{organization\}}, to her prompt template. She attaches values World Bank, UNESCO, and Badan Pusat Statistik to it.%
\footnote{Note that this part is in English now for both queries.} Re-sending queries and inspecting responses, she expands the subgroups for BPS under both her Indonesian and English response groups, such that the two subgroups are on the same screen. %
When asking for BPS data in English, both GPT-3.5 and Claude refuse to answer, whereas the same models provide BPS numbers when asked in Indonesian. Moreover, Claude's English response \emph{suggests the reader look at World Bank and UNESCO data instead,} citing those sources. \quo{That's really interesting. Wow.}

Although the study ended here, this case illustrates hypothesis iteration, limited prompts, and eagerness for cross-model comparisons, key aspects of  opportunistic exploration mode. With more time, the user might have set up an evaluation to check how models cite ``global'' sources of information when queried in English, compared to Indonesian. 

\subsection{Limited evaluation mode: Setting up an evaluation pipeline to spot-check factual accuracy.} 

How do users transition from exploratory to limited evaluation mode? %
We illustrate prototyping an evaluation and ``scaling up'' with P18, a material design student who used \oursystem{} to check an LLM's understanding conductivity values of additives to polymers. The example also depicts %
a usability issue as the user scaled up.

Like Case \#1, P18 begins in Opportunistic Exploration mode. %
They \emph{prompt / inspect / refine}---send off queries, inspect responses, revise input data or prompts. They create a prompt template with two variables: \emph{Base} and \emph{additives} (Fig.~\ref{fig:limited-eval}). Initially they start with only one Base, and four additives.  Inspecting responses, P18 is impressed with GPT-4's ability to suggest and explain \emph{specific} additives under P18's broad categories (e.g., \texttt{EMIMBF4} for \emph{Ionic Liquid}). %
They refine their questioning: %
\quo{I want to estimate the approximate conductivity value.}  %
They amend their prompt template, adding \quo{and estimate the conductivity value}. %
Reviewing responses, they find the numeric ranges roughly correct. %

They then wish to inspect the numbers in a more systematic fashion than manual inspection, and move into Limited Evaluation mode. The researcher helps P18 with how to extract the numbers, using an evaluator node, LLM Scorer, which they only saw once in the intro video. With this node, users can enter a natural language prompt to score responses. %
P18 iterates on the scorer prompt through a prompt/inspect/refine loop: first asking just for the number, %
then adding ``without units'' after they find it sometimes outputs units.\footnote{Here is where a framework like LMQL \cite{beurer2023prompting} may come in handy to improve usability of this node.} \quo{This is good. So we add some Vis Node.} They plot by \emph{additive} on the \emph{y-axis} (Fig.~\ref{fig:limited-eval}). \quo{Very good. [Researcher: Is this true?] Roughly, yes. Roughly.}

P18 then wants to ``scale up'' by adding a second polymer to their Base variable. %
They search Google for the abbreviation of a conducting polymer, Polyaniline (PANI). They paste it as a second field and re-query the prompt and scorer nodes. Skimming scores in Table Layout in two seconds: \quo{Oh, wow... %
It's really good. Because PEDOT is most [conductive].} Inspecting the Vis Node, they encounter a usability limitation: they want to \emph{group by}  \emph{Base}, when \emph{additive} is plotted in y-axis, but cannot. Plotting by \emph{Base} on the y-axis, they see via box-and-whiskers plot that PANI is collectively lower than PEDOT. They ask the researcher to export the evaluation scores.

This example illustrates limited evaluation mode, such as iterating on an evaluation pipeline (refining a scoring prompt), and beginning to ``scale up'' by extending the input data after the pipeline is set up. The user also encountered friction with usability when scaling up, wanting more options for visualization as input data complexity increased.

\subsection{Iterative Refinement mode: Tweaking an established prompt and model to attempt an optimization.}

P8 works with a German startup, and brought in a prompt engineering problem, importing a dataset and prompt template from LangChain \cite{LangChain} (\quo{we're building a custom LLM app for an e-commerce company, a virtual shop assistant}). This template had already underwent substantial revisions; thus, the participant immediately moved into iterative refinement mode, allowing us to observe interactions we could only glimpse retroactively in our interviews. 

P8's startup was using GPT-4 (because \quo{GPT-3.5 in German is really not that good}), but was curious about whether other models could perform better. He knew of Claude and PaLM2, but had been put off by needing to code up custom API calls. He also had a hypothesis that using English in  parts of his German prompt would yield better results. Upon entering the unstructured task, he imported a spreadsheet with a Tabular Data Node and pasted his three-variable prompt template in a Prompt Node, connecting them up. He then added a Python Evaluator Node to check whether the LLM stuck to a length constraint he had put in his template.  Using Grouped List layout, he compared responses between Claude and GPT-4 across ten input values for variable \emph{product\_information}. \quo{GPT4 is going over [too long]...  Claude seems to be fairly good at sticking---[opens another response group], actually, you know, we have an outlier here.} 

Looking over responses manually, he implies that he had been manually evaluating each response (prior to the study) across his ten criteria. \quo{%
I gave it... almost 10 instructions... Formal language, length, and so on. And for each... %
I now need to review it.} %
He notices that one of Claude's responses includes the word \emph{Begleiter}, a word he had explicitly instructed it to exclude: \quo{Because that was a pattern I noticed with GPT-4 that it kept using this word... %
So I'm going to try now... %
how is Claude behaving if I give this instruction in English, rather than [German]?}

To test this, he \emph{abstracts} the ``avoid the following words'' part of his prompt template into a new variable, \emph{\{avoid\_words\_  instruction\}}. He pastes the previous command into a TextFields, and add a second one---the same command but in English. %
He adds a Simple Evaluator node, checking if the response contains Begleiter. In Grouped List layout, he groups responses by \emph{avoid\_words\_instruction} and click ``Only show scores'' to only see true/false values ({\color{red} false} in red). Glancing: \quo{%
So it's not very statistically significant. But... %
GPT-4 never made the mistake, and Claude made the mistake with both English and German... %
So it doesn't matter which language... %
[Claude] will still violate the instructions.} He attaches another Simple Evaluator to test another term, remarking that in practice he would write a Python script to test all cases at once, but the study is running out of time. \quo{So Claude again violates it in both cases... [But for] English, it only violates it once---again---and in German it violates it twice. So maybe it's slowly becoming statistically significant.} As the study ends, he declares that his investigation justified his original choice: \quo{I should probably keep using GPT-4.} %

Here we see aspects of iterative refinement mode---the participant has already optimized their prompt (pipeline) and is trying to tweak the prompt and model to see if they can improve the outputs even further, \emph{according to specific criteria}. As we found in our structured task, in making decisions, users weigh trade-offs between how different models and/or prompts fulfill specific criteria, and also rank criteria importance. For P8, his ``avoid-words'' criteria seemed mission-critical, whereas word count---which he perceived Claude better at sticking to---was evidently less important.

\clearpage
\section{List of Nodes} \label{appendix:nodes}

\begin{table}[H]
    \centering
    \begin{tabularx}{\textwidth}{l|X|X}
        \toprule
        \textbf{Node Name} & \textbf{Usage} & \textbf{Special Features} \\
        \midrule
        \multicolumn{3}{c}{\textbf{Inputs}} \\
        \midrule
        TextFields Node & Specify input values to a  template variables in prompt or chat nodes. & Supports templating; can declare variables in brackets \{\} to chain inputs together. \\
        \midrule
        CSV Node & Specify input data as comma-separated values. Good for specifying many short values. & Brackets \{\} in data are escaped by default. \\
        \midrule
        Tabular Data Node & Import or create a spreadsheet of data to use as input to prompt or chat nodes. & Output values ``carry together'' when filling in multiple variables in a prompt template. Brackets \{\} in data are escaped by default. \\
        \midrule
        \multicolumn{3}{c}{\textbf{Generators}} \\
        \midrule
        Prompt Node & Prompt one or multiple LLMs. Declare template variables in \{\}s to attach input data. & Can chain together. Can set number of generations per prompt to greater than one. \\
        \midrule
        Chat Turn Node & Continue a turn of conversation with one or multiple chat LLMs. Supports templating of follow-up message. & Attach past prompt or chat output as context. Can also change what LLM(s) to use to continue conversation. \\
        \midrule
        \multicolumn{3}{c}{\textbf{Evaluators}} \\
        \midrule
        JavaScript Evaluator & Write a JavaScript function to `score' a single response. Scores annotate responses. & Boolean `false' values display in red in response inspector. \emph{console.log()} prints to node. \\
        \midrule
        Python Evaluator & Same as JavaScript Evaluator but for Python. & Can \texttt{import} packages and use \emph{print()}. \\
        \midrule
        LLM Scorer & Prompt an LLM to score responses. (GPT-4 at zero temperature is default.) & Unlike prompt chaining, this attaches the scores as annotations on existing responses. \\
        \midrule
        Simple Evaluator & Specify simple criteria to score responses as true if they meet the criteria. & Can test whether response \emph{contains}, \emph{starts with}, etc. a certain value. Can also compare against prompt variables or metavariables. \\
        \midrule
        \multicolumn{3}{c}{\textbf{Visualizers}} \\
        \midrule
        Vis Node & Plot evaluation scores. Currently only supports boolean and numeric scores. & Plots by LLM by default. Change y-axis to plot by different prompt variables. \\
        \midrule
        Inspect Node & Inspect LLM responses like the pop-up inspector, only inside a flow. & Only supports Grouped List layout. \\
        \bottomrule
    \end{tabularx}
    \parbox{\textwidth}{\caption{The nodes in \oursystem{}, grouped by type. A final node, the ``Comment Node'', allows users to write comments.} \label{tab:nodes}}
\end{table}

\end{document}